# Energy Efficient Video Fusion with Heterogeneous CPU-FPGA Devices

Jose Nunez-Yanez, Tom Sun

*Department of Electrical and Electronic Engineering, University of Bristol, BS8 1UB, UK*
*j.l.nunez-yanez@bristol.ac.uk, tom.sun@bristol.ac.uk*

***Abstract*—This paper presents a complete video fusion system with hardware acceleration and investigates the energy trade-offs between computing in the CPU or the FPGA device. The video fusion application is based on the Dual-Tree Complex Wavelet Transforms (DT-CWT). Video fusion combines information from different spectral bands into a single representation and advanced algorithms based on wavelet transforms are compute and energy intensive. In this work the transforms are mapped to a hardware accelerator using high-level synthesis tools for the FPGA and also vectorized code for the single instruction multiple data (SIMD) engine available in the CPU. The accelerated system reduces computation time and energy by a factor of 2. Moreover, the results show a key finding that the FPGA is not always the best choice for acceleration, and the SIMD engine should be selected when the wavelet decomposition reduces the frame size below a certain threshold. This dependency on workload size means that an adaptive system that intelligently selects between the SIMD engine and the FPGA achieves the most energy and performance efficiency point.**

***Keywords—video fusion; Energy efficient; Hybrid FPGA***

## I. INTRODUCTION

Multi-sensor video data with visible and infrared images is increasingly being utilized in applications such as medical imaging, remote sensing and security applications. Multi-sensor data presents complementary information about the region surveyed and fusion provides an efficient method to combine the complementary information for better data analysis. Video fusion is just a special case of image fusion when two or more frames of different video sources are fused together continuously into a single fused video. Image fusion can be performed at signal, pixel, feature and symbolic levels, and this paper focuses on the pixel level algorithms presented in [1] based on wavelet transform techniques [2]. Compare to other schemes [3], wavelet transform achieves better signal to noise ratios and improved perception with no blocking artefacts. Moreover, among all the wavelet transform that applied to multifocal, remote sensing and medical image fusion, the use of the *Dual-Tree Complex Wavelet Transform* (DT-CWT) has been shown to produce significant fusion quality improvement [4]. The algorithm used in this paper consists in applying DT-CWT to infrared and visible frames, combining the obtaining coefficients using a fusion rule and then proceeding to perform the inverse DT-CWT for reconstruction.

The proposed system is based on the ZYNQ System-on-Chip and the CPU and the FPGA work together to run the algorithm. The whole system runs under the Linux OS with a customized kernel level Linux driver. The main contributions of this paper are:

1. We create an open-source complete fusion system including processing engine, drivers, hardware interfaces and cameras. The most compute intensive parts of the algorithm are accelerated based on HLS tools using the FPGA and vectorized based on SIMD functions using the NEON engine.
2. We demonstrate the performance and energy advantages of using a heterogeneous platform for video fusion comparing to a software-only solution.
3. We show that depending on the amount of data and frame size the most efficient way to compute the wavelet transforms changes between FPGA and CPU so an adaptive solution that selects the optimal hardware at run-time is preferred.

The remaining of this paper is organized as follows. Section II lists related work in this research area. Section III provides some basic knowledge of the DT-CWT based fusion algorithms and Section IV discusses optimization for SIMD execution. Section V introduces our hardware architecture to implement the DT-CWT with a customized kernel level Linux driver, followed by Section VI, which presents our system architecture to capture and fuse multi-sensor data. Section VII compares the performance and power consumption under the ARM CPU, NEON SIMD and FPGA configurations and Section VIII concludes the paper.

## II. RELATED WORK

Previous research on FPGA-based fusion systems is available in recent literature. Jasiunas et al. [5] presented a wavelet based image fusion system for unmanned airborne vehicles. This is a very early attempt to develop image fusion systems on reconfigurable platform alone that achieved latency of 3.81 ms/frame for visible and infrared 8-bit images of 512x512 pixel resolution. Sims and Irvine [6] presented an FPGA implementation of pyramidal decomposition based video stream fusion. This framework can achieve a 30 frame/s, real-time fuse of video streams in grayscale video graphic arrays (VGA). Yunsheng et al. [7] presents a real-time image processing system to combine the video outputs of an uncooled infrared imaging system and a low-level-light TV system. Song et al. [8] proposed an image fusion implementation based on Laplacian pyramid decomposition of two-channel VGA video for a better fusion quality and reasonable frame rate of 25 frame/s. Mohamed and El-Den [9] applied five different measures to evaluate the performance of several different fusion techniques and the hardware implementation of DCT, DWT and PCNN-based fusion algorithms are studied. However, although these designs achieves performance enhancement to do image fusion on FPGA, the fusion algorithms they used are not state-of-the-art.

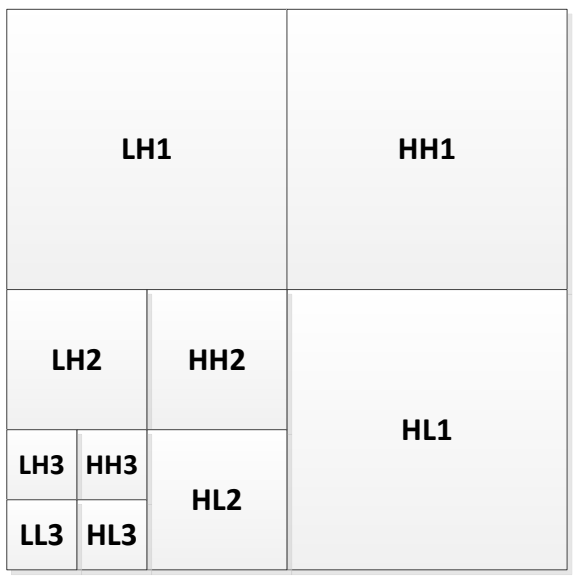

Fig. 1 Two dimensional Discrete Wavelet Transform

Tao et al. [10] proposed an image enhancement and fusion system to improve visibility. In this paper, two videos are captured by CCD and LWIR cameras and fused by implementing DT-CWT fusion algorithms in Xilinx Virtex-II environment. Gudis et al. [11] built an embedded vision service framework on ZYNQ SoC with a "plug-and-play" capability to allow the service-based software to take advantage of the hardware acceleration blocks available and perform the remainder of the processing in software. These designs share some similarities with our system but focus on the fusion quality more than the performance and energy efficiency.

## III. The DT-CWT Based Fusion Algorithm

The aim of the wavelet transformation is to represent signals using a superposition of wavelets. The *Discrete Wavelet Transform* (DWT) is a spatial-frequency decomposition of a signal, which ensures the signal being decomposed into normalized wavelets at octave scales [12]. When applied to two-dimensions, signals are separately filtered and down-sampled in the horizontal and vertical directions. This creates four sub-bands at each scale, namely high-high (HH), high-low (HL), low-high (LH) and low-low (LL), as shown in Fig. 1. The name of each sub-band denotes the horizontal frequency first and then the vertical frequency. A multi-resolution decomposition of image can then be achieved by recursively applying filtering to the low-low sub-band. The number shown in Fig. 1 denotes the decomposition level and it can be seen that larger decomposition levels indicate a reduction in frame size. This feature will have implications on the preferred compute engine as will be explained in Section VII.

The DT-CWT transforms signals use two separate DWTs and apply spatial filters recursively to create frequency sub-bands. The application of DT-CWT to 2-D image is achieved by separable complex filtering in two dimensions. The DT-CWT is able to distinguish between positive and negative orientations and divides the horizontal and vertical sub-bands into six distinct sub-bands at each scale with the orientations of ±15°, ±45° and ±75°. Moreover, the DT-CWT gives perfect reconstruction due to the biorthogonal nature of the filters and also delivers approximate shift-invariance.

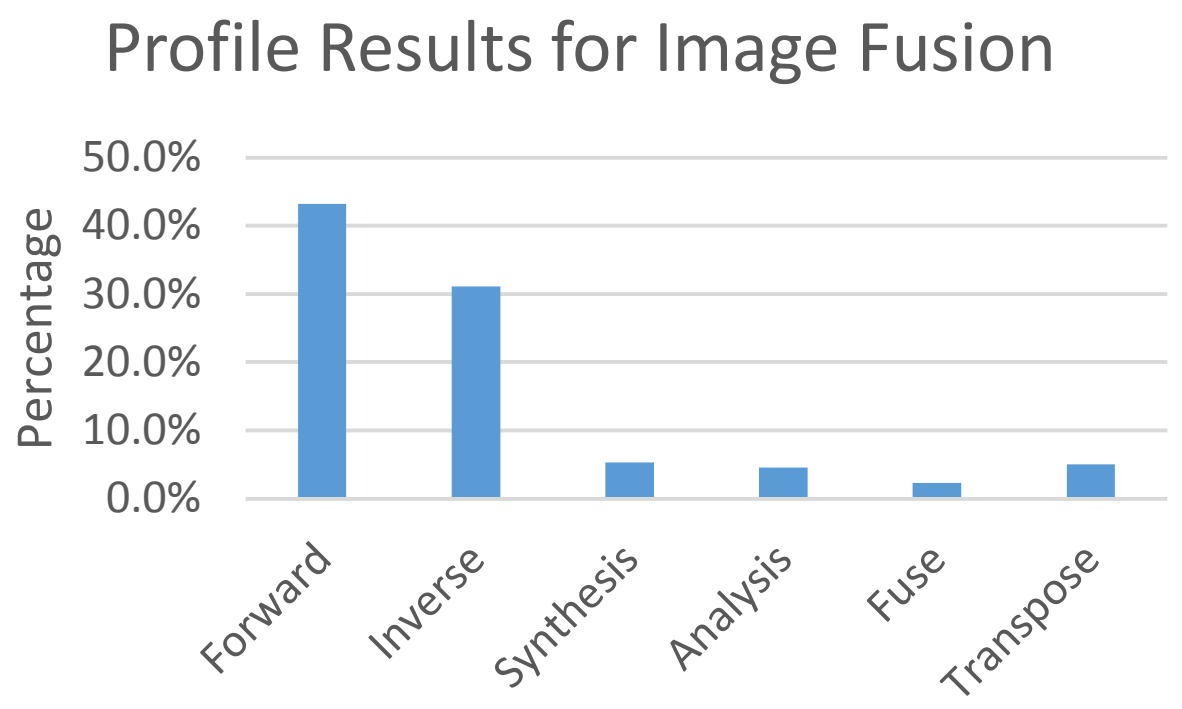

Fig. 2 Profiling results of fusing two input images

In this paper, the whole fusion algorithm with the forward and inverse DT-CWTs is written in C++ and executed by the ARM Cortex A9 Processor. The profiling results of the fusion process, as shown in Fig. 2, indicate that the forward and inverse DT-CWT are the most compute- and energy intensive tasks. Therefore, these parts of the algorithm are the ones selected for acceleration. The ZYNQ device designed by Xilinx offers two alternatives for code acceleration, either using the FPGA that can be made cache coherent with the CPU thanks to the Acceleration Coherence Port (ACP) or using the NEON SIMD engine that is a part of the ARM Cortex A9 CPU. The next two sections describe how each of these methods is deployed.

## IV. SIMD Acceleration

NEON is a 128-bit SIMD architecture extension for the ARM Cortex-A series processors, designed to load, compute and store data using vector registers so that multiple, independent data can be processed concurrently. It has 32 registers and each of them is 64-bit wide, which can also be treated as 16 registers, each with a width of 128-bit. Given the nature of recursive application of spatial filters in the forward and inverse DT-CWT with no loop-carry dependency, there are opportunities to optimize these parts of the codes using SIMD functions in order to exploit the embedded NEON engine. In this paper, vectorization was attempted both at the programmer level, by manually using various NEON intrinsics defined in the *arm_neon.h* header file and at the compiler level, by inserting "*-mfpu=neon -ftree-vectorize*" while compiling using g++ for auto-vectorization. Fig. 3 shows the extraction of both the automatic and manual vectorization of one function in a *for-loop*. To enable the NEON auto-vectorization, all pointers were declared using the "*_restrict*" keyword to inform the compiler that the location accessed through a specific pointer was not to be accessed through any other pointer within the current scope. The fixed loop length L was a multiple of 4 and has its bottom two bits masked, so that the compiler can perform otherwise unsafe vectorizations. For manual vectorization, the 128-bit

```
/*Original Code Before Vectorization*/
...
for (int k = 0; k < L ; k++){
    (*hpCoeff) = (*hpCoeff) + hp2[k]*y2[k];
}
...

/*Use of NEON Auto-vectorization*/
...
for (k = 0; k < (L & ~3); k += 4){
    (*hpCoeff) = (*hpCoeff) + hp2[k]*y2[k];
}

/*Use of NEON intrinsics after Vectorization*/
...
float32x4_t hp2_q, y2_q;
float32x4_t hpCoeff_q, lpCoeff_q;
float32x2_t tmphp[2];

for (k = 0; k < (L & ~3); k += 4){
    hp2_q = vld1q_f32(&hp2[k]);
    y2_q  = vld1q_f32(&y2[k]);
    hpCoeff_q = vld1q_f32(&hpCoeff[k]);
    hpCoeff_q = vmlaq_f32(hpCoeff_q, hp2_q, y2_q);
    }
tmphp[0] = vget_high_f32 (hpCoeff_q);
tmphp[1] = vget_low_f32 (hpCoeff_q);
tmphp[0] = vpadd_f32 (tmphp[0], tmphp[1]);
tmphp[0] = vpadd_f32 (tmphp[0], tmphp[0]);
(*hpCoeff) = vget_lane_f32 (tmphp[0], 0);
...
```

Fig. 3 Sample code Extraction for SIMD vectorization

vector registers were used, each declared by the NEON intrinsic *"float32x4_t"*, to store four floating point numbers into a single register. After adding and multiplying in vector form, the four floating point numbers residing in the 128-bit register added with each other in order to return a single 32-bit floating point number.The loop number should be fixed at the multiple of the number of lanes in the vector register. Otherwise, extra steps, used to handle the remaining loop iterations in scalar form, will cause performance degradation. In our paper where the NEON quad-word registers were used to store data with type of 32-bit float, an iteration count with a multiple of 4 is used. Both the manual and auto vectorization produced the similar performance enhancement, and the results are presented in Section VII.

## V. FPGA Acceleration

To achieve the FPGA acceleration, the forward and inverse DT-CWT were mapped to the PL (FPGA) side of ZYNQ to create a hardware wavelet engine controlled by the PS (CPU) side. This means that the input images are decomposed and reconstructed in hardware. The hardware accelerator has been created using the VIVADO_HLS high-level synthesis tools increasing productivity compared with a traditional RTL design. The ZYNQ Processing System and Programmable Logic (PS-PL) interface is created to transfer commands, filtered coefficients, transformed coefficients and pixel data between the PS and the PL. The general purpose 32-bit ports do not obtain the require performance and every transfer requires around 25 clock cycles with the CPU moving the data itself. For this reason we created a custom DMA engine using the synthesis support of *memcpy* by VIVADO_HLS. Cache coherence is ensured by using the Accelerated Coherence Port (ACP) to connect the PL to the PS. The code for VIVADO_HLS is configured to generate two interfaces. An AXI4Lite slave interface is used to load filter coefficients and send commands to the engine to enable the execution of the forward and inverse transform. An AXI4M interface is used to load and store pixel and transformed data using the hardware implemented *memcpy* function through the ACP port. Fig.4 shows a section of the code corresponding to the forward wavelet transform synthesized into FPGA logic and memory by the VIVADO_HLS tools with full code available in [13].

```
//read data
memcpy(buff_in, (float *)(memory + in_offset), (outwidth * 2 + 12)*sizeof(float));

wav_engine_master_label0:for (int i = 0; i<(outwidth + 6); i++)
{

       input_a = (data_t)buff_in[i * 2];
       input_b = (data_t)buff_in[i * 2 + 1];

       hpMult = coeff_register_hp[0] * shift_register[0];
       lpMult = coeff_register_lp[0] * shift_register[0];
       hpAcc = hpMult;
       lpAcc = lpMult;

       wav_engine_master_label1:for (int j = 1; j < 11; j++)
       {
              lpMult = coeff_register_lp[j] * shift_register[j];
              hpMult = coeff_register_hp[j] * shift_register[j];
              hpAcc += hpMult;
              lpAcc += lpMult;
              shift_register[j - 1] = shift_register[j + 1];
       }
       lpMult = coeff_register_lp[11] * shift_register[11];
       hpMult = coeff_register_hp[11] * shift_register[11];
       hpAcc += hpMult;
       lpAcc += lpMult;
       shift_register[10] = input_a;
       shift_register[11] = input_b;
       if (i > 5)
       {
                buff_out[i * 2 - 12] = (float)hpAcc;
                buff_out[i * 2 + 1 - 12] = (float)lpAcc;
       }
}
//write data
memcpy((float *)(memory + out_offset), buff_out, (outwidth * 2)*sizeof(float));
```

Fig. 4 Sample code Extraction for FPGA synthesis

The *memcpy's* move data between the external DDR memories and internal BRAMs and the *for* loops create the filters with the help of an internal shift register. The final *if* makes sure that only the correct outputs are written to the output buffers. Additional pragmas are used to ensure that the tool adds the require AXI interfaces and pipeline registers to obtain an initialization interval of one clock cycle so a new input enters the pipeline in each clock cycle. Notice that the current VIVADO_HLS tools do not pipeline the *memcpy's* that need to complete before the loop processing can start. It is important to note that all the logic required to implement these functions is created on the PL side by VIVADO_HLS. Control variables not shown in this sample code activate one of three possible modes that correspond to 1) filter coefficient loading, 2) forward transforms and 3) inverse transform. The PL works with a single clock frequency of 100 MHz to meet all the timing constraints while the PS works at the default of 533 MHz. With this setup, we wrote a kernel level Linux driver to allocate memory that can be accessed by the accelerator with physical addresses and by the processor with virtual addresses. The driver uses the standard *"memcpy"* function, implemented in this case in software at the user level, for data transfer. For this to work, it is necessary to obtain the physical addresses at which the memory is created by the *"kmalloc"* calls in the kernel driver, and then use the memory-map calls *"mmap"* to obtain remapped virtual addresses in user space that can be used by standard *"memcpy"*. Additionally, the Linux driver implements the *"ioctl"* function, which can be used to control how the data movements take place. In our case, we used this to create different read and write offsets to the kernel allocated memory. To increase the performance of the system we divided the kernel memory into two areas or buffers. This double buffering mechanism is used to parallelize the transfer and processing of data from user space to kernel space as illustrated in Fig. 5. This approach reduces latency and hardware complexity compared with buffering the whole image in the FPGA memory. The input and output buffers have a size of 4096 32-bit, divided into two areas of 2048 32-bit, which is suitable for an image width up to 2048 pixels. Table I shows the implementation complexity of this hardware wavelet engine.

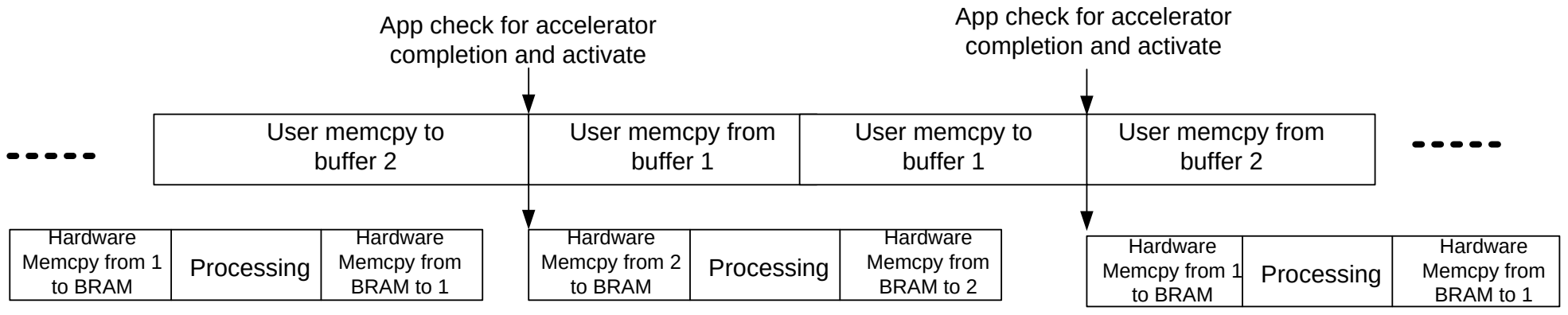

Fig.5 Design of the Kernel Level Linux Driver

TABLE I. IMPLEMENTATION COMPLEXITY OF WAVELET ENGINE

| **Wavelet Engine** | **Implementation Complexity Part:** ***xc7z020clg484-1*** | | |
|---|---|---|---|
| | ***Unitization*** | ***Available*** | ***Percentage*** |
| Registers | 23412 | 106400 | 22% |
| LUTs | 17405 | 53200 | 32% |
| Slices | 7890 | 13300 | 59% |
| BUFG | 3 | 32 | 9% |

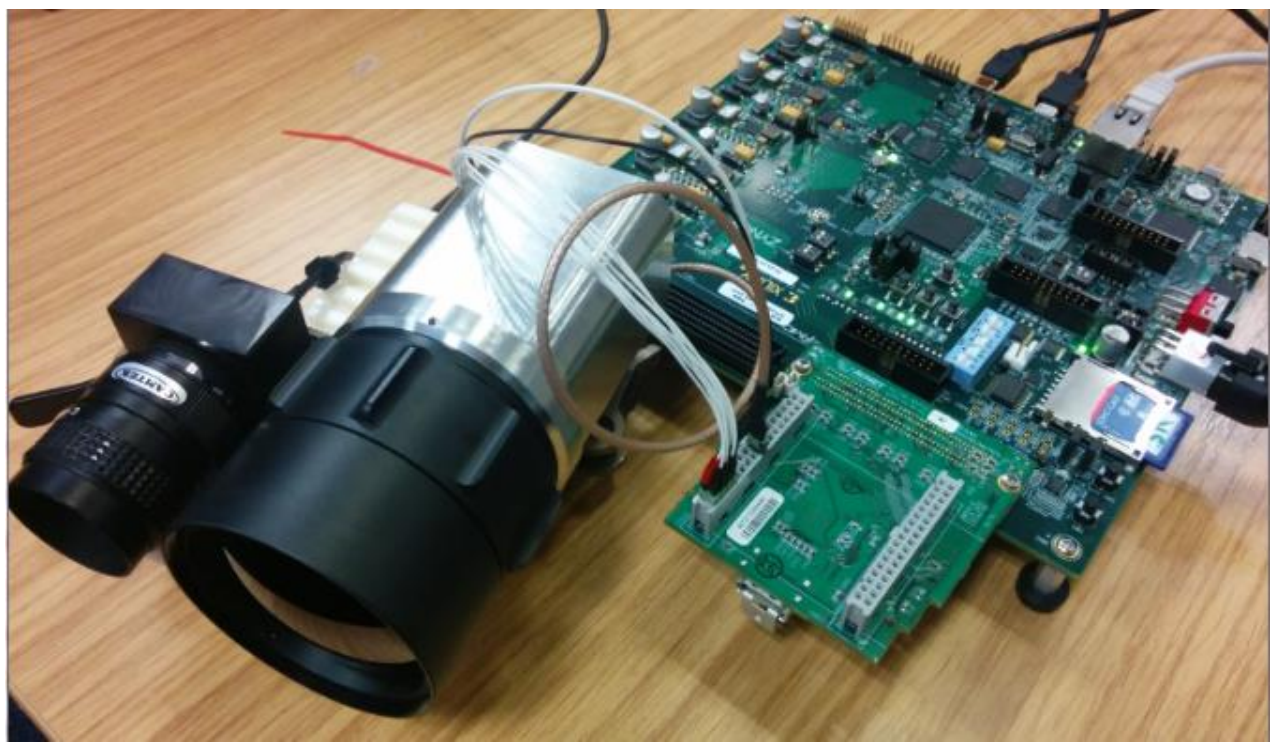

Fig. 6 System prototype

## VI. THE SYSTEM ARCHITECTURE

This section describes the overall system architecture we implemented to capture and fuse the multi-sensor data. In this paper, we have used the ZYNQ-based ZC702 Evaluation Board running UBUNTU Linux OS. A web camera and a thermal camera were placed together to capture the same scene before fusion. The real system and the overall architecture are shown in Fig. 6 and Fig. 7 respectively. As shown in Fig. 7, the input video captured by the web-camera (Logiteck webcam C160) is decoded on the PS side through the USB-OTG port, and the video captured by the thermal-camera (Thermoteknix MicroCAM 384H XTi) is decoded by a customized BT656 decoder system implemented on the PL side, through one of the FPGA Mezzazine Card (FMC) connectors. According to Fig. 7, the input thermal pixel data is decoded by the BT656 decoder

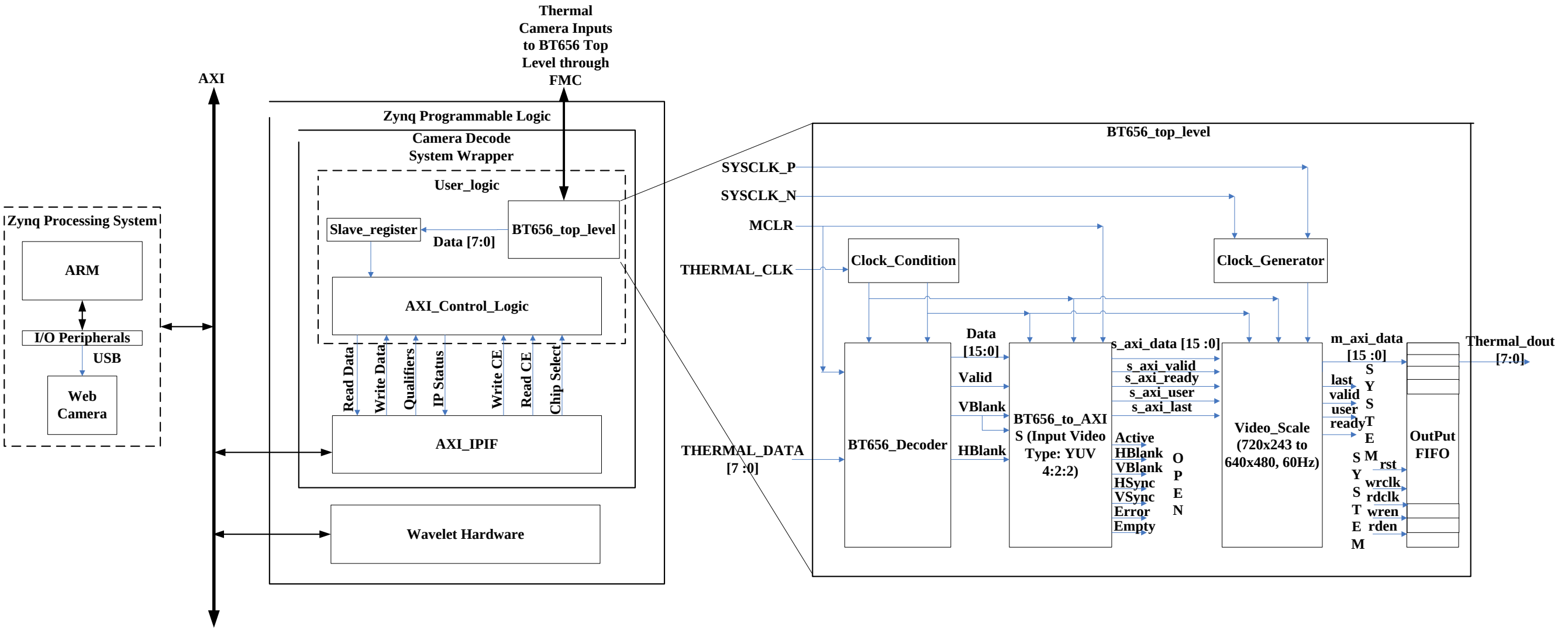

Fig. 7 Overview of the system Design and the BT656 Decoder

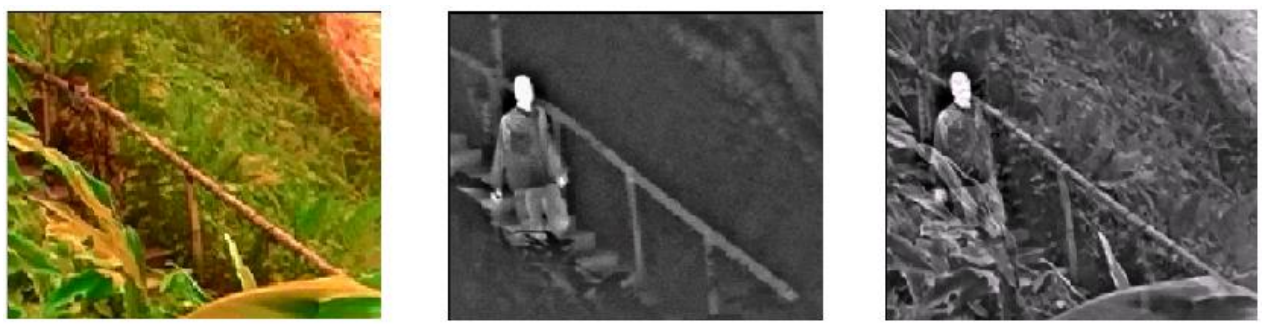

7(a) Web-cam frame 7(b) Thermal-cam frame 7(c) fused frame and sent for scaling through an AXI interface. The whole frame of the video is then stored in the output FIFO, waiting to be taken for decomposition. The AXI control signals guarantee that a new frame will be stored in the output FIFO only after the previous frame is taken by the wave engine hardware. The *clock_condition* component uses the clock signal (*thermal_clk)* from the thermal camera to drive the BT656_Decoder, and the *clock_generator* component uses the differential clock from the system to drive the *video_scale* component. The data transfer between the PS and the PL is done through the AXI interface. Both input videos are decoded into continuous pixel frames and sent to the wavelet hardware on the PL side for DT-CWT decomposition. The transformed coefficients are sent to the PS for fusion and then sent back to the wavelet hardware for inverse DT-CWT reconstruction. Since the whole system is running under Linux OS, the decoded and the fused videos are shown on screen using OpenCV funtions, with no external video connectors or cables required. Fig. 8 demonstrates the video frame captured by the web-camera and the thermal-camera and the fused frame of the two. The original video captured by the web-camera was gray-scaled before fusing. The full demonstration of the video capturing and fusion is available at [13]. To ensure reproducible research we have also made a demonstration/verification system with source code for the ZC702 board with download details available in [13].

Fig. 8 Demonstration of the Designed Fusion System

## VII. Results Analysis and Comparison

This section compares the fusion performance and power consumption when the forward and inverse DT-CWTs are executed by the ARM processor, the NEON engine and the FPGA respectively. The designed system input videos with frame size of 88x72 pixels and output a fused video with the same frame size. The small frames are selected due to constraints of the longwave infrared sensor which are much more limited in resolution compared with standard camera sensors (i.e. Lepton module at 80x60 pixels). Wavelet processing involves a number of decomposition levels that reduce the size of the frame each time. In this test the decomposition level of the CT-DWT was varied and four sets of smaller frames were also extracted from the original input frames for fusion with a smaller frame size. The performance comparisons of each frame size are shown in Fig. 9. The results were obtained by profiling when 10 input frames were decomposed, fused and reconstructed continuously. Compared to the situation when the forward DT-CWT was executed by the ARM processor, Fig. 9(a) shows a performance enhancement (defined by the reduction of the execution time) of 55.6% when using the FPGA and a performance enhancement of 10% when using the NEON engine to forward transform the full frames (88x72 pixels). However, for smaller extractions of the full frame at 32x24 pixels, execution of the forward DT-CWT by FPGA caused a 36.4% performance degradation (defined by the increment of the execution time) compared to the situation when the forward DT-CWT was executed by the NEON engine. The forward transform using FPGA at this point took longer than that using the ARM processor since the overhead of passing commands from the PS to the PL is relatively significant at this level. As the frame size increases, the advantage of using FPGA outperforms that of using the NEON engine with the breaking point at frame size between 35x35 and 40x40 pixels. Similar situations happened for the inverse DT-CWT transform, as shown in Fig. 9(c). Compared to the situation when the inverse DT-CWT was executed by the ARM processor, execution using the FPGA to transform the full frame size (88x72 pixels) provided 60.6% performance enhancement while the execution using the NEON engine provided 16% performance enhancement. The FPGA still provided worse performance than the NEON engine at frame size 35x35 and 32x24 pixels, and it only outperformed the NEON engine when the frame size increased past 40x40 pixels. Concerning the total time taken to decompose, fuse and reconstruct the 10 consecutive frames, Fig. 9(b) shows the same trends as described above. The ARM+FPGA execution outperformed the ARM+NEON only when the frame size was increased beyond 40x40 pixels. At full frame size (88x72 pixels), the FPGA provided 48.1% performance enhancement and the NEON engine provided 8% performance enhancement. The energy consumption of fusion at each frame size was calculated using the power values, measured by power-recording software running simultaneously with the fusion process, and the total time taken shown in Fig. 9(b). Fusing using only the ARM processor consumes approximately the same power as using ARM+NEON. However, fusing using ARM+FPGA consumes 3.6% more power (19.2mW) due to the extra power introduced by the wave engine hardware in the PL side. This is a net power increment considering both the power decreased on the PS side due to the reduced processor load and the power increased on the PL side due to the extra hardware activated. Fig. 10 shows the total energy comparison when 10 frames with different sizes were decomposed, fused and reconstructed continuously. Compare to the ARM only situation, ARM+FPGA saves 46.3% of total energy consumption when fusing images with full frame size, while ARM+NEON saves 8% of it. The use of ARM+FPGA is only more energy efficient than the use of ARM+NEON when the frame size is larger than 40x40 pixels. The breaking point exists at the frame size between 40x40 and 64x48 pixels, and starting from the breaking point, the larger the frame size to be fused, the more energy efficient is the ARM+FPGA processing mode compared to both ARM only and ARM+NEON processing mode. For larger frames it is clear that the performance and energy advantages of the FPGA device are obvious but in certain constrained scenarios a run-time selection of the accelerator can be optimal.

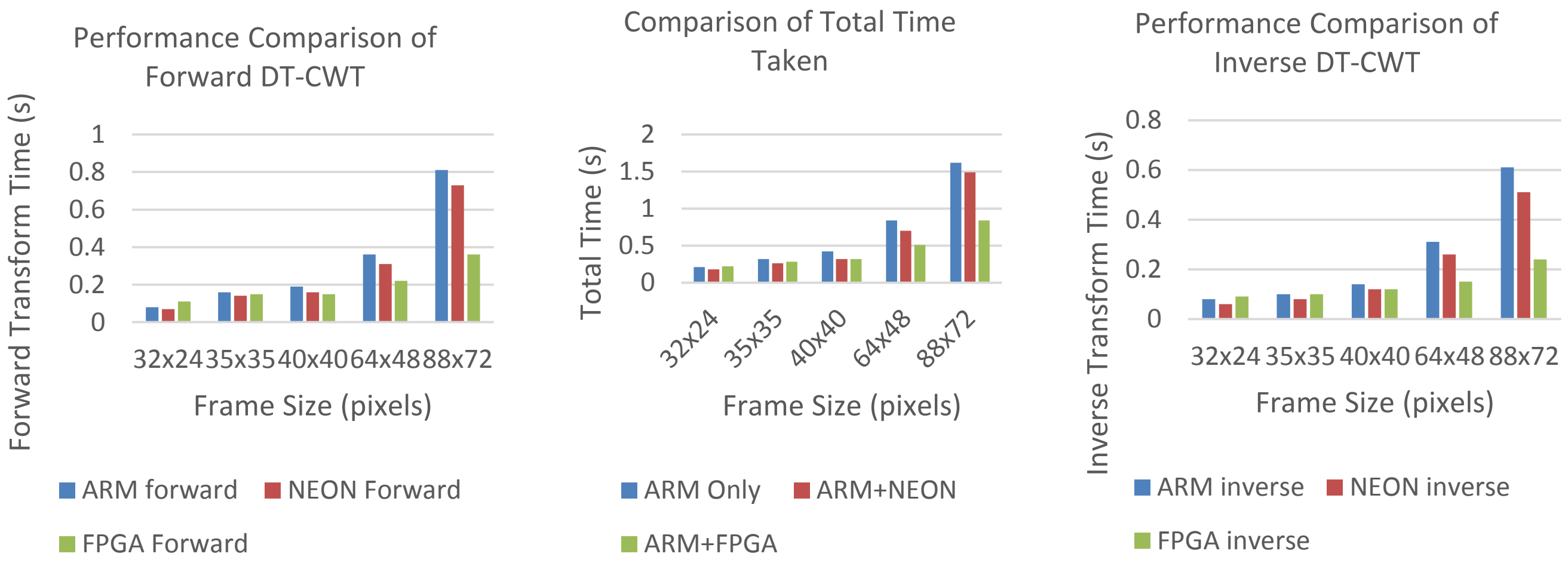

(a). Comparison of Forward DT-CWT (b). Total Time Comparison (c). Comparison of Inverse DT-CWT

Fig. 9 Performance Comparison when the Forward and Inverse DT-CWT are executed by ARM, NEON and FPGA

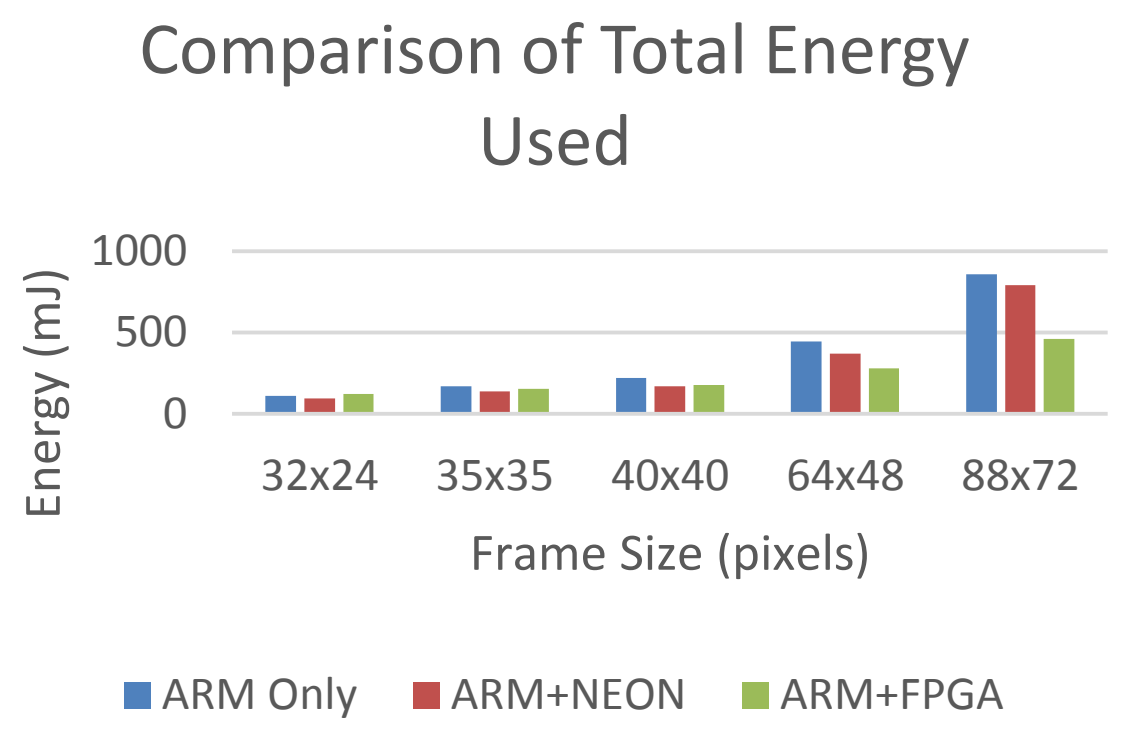

Fig. 10 Total Energy Comparison of different execution mode

## VIII. Conclusions and Future Work

This paper has presented an energy efficient video fusion design which can capture both visible and infrared videos simultaneously and fuse them by applying a fusion algorithm based on the DT-CWT. In our design, the most compute intensive tasks, namely the forward and inverse DT-CWT were vectorized to exploit the NEON SIMD functionalities and mapped to a closely coupled FPGA with a customized Linux kernel level driver to release the processor load. The performance and energy consumption of fusing input frames with different levels of decomposition was compared considering configurations when the fusion process was executed by ARM processor only, ARM with NEON engine and ARM with FPGA accelerators. Comparing to the execution using the ARM processor only, using the FPGA can save 55.6% (vs 10% using NEON) and 60.6% (vs 16% using NEON) of the execution time for the forward and inverse DT-CWT execution respectively at the frame size of 88x72 pixels. The experiments also show that the FPGA is not always the best choice and the NEON engine should be selected when the wavelet decomposition reduces the frame size below a certain threshold. In essence, using the FPGA generates overheads when preparing and transferring data and results between PL and PS sides over the AXI interconnect, which could be counter-productive if the workload is small. This dependency on workload size means that an adaptive system that intelligently selects between the NEON engine and the FPGA should achieve the most energy and performance efficient point. Future work will involve extending this design to make the system automatically choose the resources (NEON or FPGA) to execute when fusing with different frame sizes and decomposition levels.

## References

[1] D.K. Sahu and M.P. Parsai, “Different Image Fusion Techniques – A Critical Review,” Int’l Journal, Modern Engineering Research (IJMER), vol. 2, Issue. 5, Sept. 2012, pp. 4298-4301.

[2] S. Nikolov, P. Hill, D. R. Bull, and C. N. Canagarajah, "Wavelets for image fusion," in Wavelets in Signal and Image Analysis: From Theory To Practice, Kluwer Academic Publishers, 2001.

[3] A. Toet, “Hierarchical Image Fusion,” Machine Vision and Applications, March 1990, pp. 1-11

[4] Remove for blind review

[5] D. Jasiunas et al., “Image fusion for uninhabited airborne vehicles,” Proc. Int’l. Conf. FPT, Dec 2002, pp. 348-351.

[6] O. Sim and J. Ivine, “An FPGA implementation of pattern-selective pyramidal image fusion,” in Proc. Int. Conf. FPL, 2006, pp. 1-4.

[7] Q. Yunsheng et al., “The real-time processing system of infrared and LLL image fusion,” Proc. Int’l. Symp. Photoelectron Detection Image process, 2008, pp. 66231Y-1- 66231Y-9.

[8] Y. Song et al., Implementation of Real-time Laplacian Pyramid Image Fusion Processing based on FPGA,” Proc. SPIE, vol. 6833, 2007, pp. 16-18.

[9] M.A. Mohamed and B.M. El-Den, “Implementation of Image Fusion Techniques Using FPGA,” Int’l. Journal of Computer Science and Network Security, vol.10, No.5, 2010, pp. 95-102.

[10] L. Tao et.al., “A Multi-sensor Image Fusion and Enhancement System for Assisting Drivers in Poor Lighting Conditions,” Proc. Applied Imagery and Pattern Recognition Workshop, 2005, pp. 1-6.

[11] E. Gudis et al., “An Embedded Vision Services framework for Heterogeneous Accelerators,” Proc. Computer Vision and Pattern Recognition Workshops 2013, pp. 598-603.

[12] R.P. Singh, R.D. Dwivedi, and S. Negi, “Comparative Evaluation of DWT and DT-CWT for Image Fusion and De-noising,” Int’l. Journal, Applied Information Systems (IJAIS), vol. 4, Sept. 2012, pp. 40-45.